\documentclass[fleqn]{2023SCGE}
\usepackage{aas_macros}
\usepackage[sorting=none]{biblatex}
\usepackage{multicol}

\usepackage{hyperref}%PDF??????
\begin{document}

\ensubject{subject}

%%%%%%%%%%%%%%%%%%%%%%%%%%%%%%%%%%%%%%%%%%%%%%%%%%%%%%%
%%% Authors do not modify the information below
%%% ????????????????
%%% ??????????, ????????????{}, ???????????????????
%Letter to the Editor??Article%??????
\ArticleType{Article}%??Article
\SpecialTopic{SPECIAL TOPIC: }%???????
\Year{2025}
\Month{January}
\Vol{??}
\No{?}
\DOI{??}
\ArtNo{000000}
\ReceiveDate{January XX, 2025}
\AcceptDate{January XX, 2025}
%\OnlineDate{January 1, 2016}
%%%%%%%%%%%%%%%%%%%%%%%%%%%%%%%%%%%%%%%%%%%%%%%%%%%%%%%

%%% title: ????
%%%   \title{title}{title for citation}
\title{Forecast of gravitationally lensed Type Ia supernovae time delay measurement by Muztage-Ata 1.93m Synergy Telescope}{Forecast of gravitationally lensed Type Ia supernovae time delay measurement by Muztage-Ata 1.93m Synergy Telescope}

%%% Corresponding author: ???????
%%%   \author[number]{Full name}{{email@xxx.com}}
%%% General author: ???????
%%%   \author[number]{Full name}{}
\author[1,2]{Guanhua Rui}{}%
\author[3]{Wenwen Zheng}{}
\author[3]{Zizhao He}{}%\protect\\?§Þ??§Ø????
\author[3]{\\Yiping Shu}{}%
\author[4]{Xinzhong Er}{}
\author[3]{Guoliang Li}{}
\author[1,2]{Bin Hu}{}

%%% Author information for page head. ?¨¹?§Ö????????
%%% ??????????????, ??????????author
\AuthorMark{Guanhua Rui}%\authorcr????????

%%% Authors for citation. ????????§Ö????????
%%% ??????????????, ??????????author???
\AuthorCitation{Guanhua Rui et al}

%%% Address. 
%%%   \address[number]{Address, City {\rm Postcode}, Country}
\address[1]{Institute for Frontier in Astronomy and Astrophysics, Beijing Normal University, Beijing, 102206, China;}
\address[2]{School of Physics and Astronomy, Beijing Normal University, Beijing 100875, China;}
\address[3]{Purple Mountain Observatory, Chinese Academy of Sciences, Nanjing, Jiangsu, 210023, China;}
\address[4]{Tianjin Astrophysics Center, Tianjin Normal University, Tianjin 300387, P.R.China}

%\contributions{}%

%%% Abstract. 
\abstract{
Strong lensing time delay measurement is a promising method to address the Hubble tension, offering a completely independent approach compared to both the cosmic microwave background analysis and the local distance ladder. As a third-party examination of the Hubble tension, this method provides a unique perspective. Strongly lensed quasar (glQSO) systems have demonstrated significant potential in tackling this issue, achieving an impressive \(2\%\) accuracy level. However, advancing to \(1\%\) or sub-percent accuracy is challenging due to several intrinsic limitations of glQSOs.
Fortunately, strongly lensed supernovae (glSNe) offer a more robust solution, thanks to their characteristic light curve, significant brightness variations, and additional advantages. The Muztagh-Ata 1.93m Synergy Telescope (MOST) is an exceptional instrument for monitoring strong lensing time delays. In this study, we simulate the follow-up multi-band light curve monitoring for glSNe Ia systems, which are expected to be firstly discovered by the Chinese Survey Space Telescope (CSST). Our results show that with \(300s \times 9\) exposures in each epoch, MOST can achieve a signal-to-noise ratio (SNR) of approximately 50 for the brightest images of glSNe Ia, while even the faintest images maintain an SNR of at least 7. Using a standard SNe Ia light curve template for fitting, we measured the time delays. With a 2-day cadence, MOST achieves a time delay error of only a few hours, with the bias typically remaining below one hour. 
This study highlights the capability of MOST to significantly advance the precision of time delay measurements, offering a promising path toward resolving the Hubble tension.
}



\maketitle

%\tableofcontents%?????

%%%%%%%%%%%%%%%%%%%%%%%%%%%%%%%%%%%%%%%%%%%%%%%%%%%%%%%
%%% The main text. ???????
%???????????????????\cref{fig1}
%\twocolumn%\onecolumn
%%%%%%%%%%%%%%%%%%%%%%%%%%%%%%%%%%%%%%%%%%%%%%%%%%%%%%%
\begin{multicols}{2}

%\Authorfootnote

%%%%%%%%%%%%%%%%%%%%%%%%%%%%%%%%%%%%%%%%%%%%%%%%%%%%%%%
\section{Introduction}\label{section1}
The Hubble constant (\(H_0\)) determined from cosmic distance ladder measurements \cite{Riess2022} exhibits a discrepancy larger than \(5\sigma\) when compared to the value inferred within the \(\Lambda\)CDM framework based on cosmic microwave background (CMB) observations \cite{Planck2020}. This discrepancy highlights the need for a third method that can independently and accurately measure \(H_0\), to determine whether the deviation arises from unknown systematic errors or new physics. As early as half a century ago \cite{Refsdal1964}, it was recognized that measuring time delays between multiple images in strong gravitational lens systems provides an independent approach to determining cosmological parameters.

The fundamental principle involves the gravitational lensing effect, where a massive celestial object, such as a galaxy or galaxy cluster, located between an observer and a background source, acts as a lens, distorting the light emitted by the background source and creating multiple images on the lens plane. Due to differences in light paths, these multiple images reach the observer at different times, resulting in time delays. For galaxy-scale lens systems, these time delays typically range from days to months, whereas for galaxy-cluster-scale lens systems they can extend to several years. By modeling the mass distribution of the lens system, inferring the lensing potential at the positions of the corresponding image points, correcting for external shear based on the shapes of surrounding galaxies, and subsequently calculating the time delays by comparing the light-curve variations of the multiple images of the background source, we can deduce the cosmological parameters associated with the observed time delays.

In recent years, substantial progress has been made in using time delays from gravitational lens systems to constrain \(H_0\). For instance, the \texttt{H0LiCOW} project \cite{Suyu2017, Wong2020}, which employed high-quality imaging data and analyzed lensed quasar systems, demonstrated the effectiveness of this approach and provided a precise measurement of \(H_0\). Their results are in strong agreement with \(H_0\) values obtained from cepheid variable stars \cite{Riess2016, Riess2019}, but significantly exceed the value derived from the Planck Collaboration's CMB measurements \cite{Planck2020} by approximately 3\(\sigma\).

For gravitationally lensed quasars (glQSO), several significant challenges need to be addressed. Initially, the light curve variation in QSOs is minimal, as approximately 70\% of known QSOs exhibit variability at \(\sigma = 0.1 \, \text{mag}\) (around 30\% at \(\sigma = 0.2 \, \text{mag}\)).\cite{Hartwick1990} Microlensing effects pose contamination, necessitating high-cadence observations for accurate time delay estimation. Secondly, the inherent complexity of quasars results in the stochastic nature of their light curves, as noted in previous studies \cite{Liao2015}. Thirdly, quasars tend to have much higher luminosities compared to their host galaxies, making the lensing modeling process through direct observations of glQSO systems significantly more challenging \cite{Suyu2017, Wong2020, Birrer2019}. This luminosity contrast also introduces a selection effect, where confirmed glQSO systems tend to have larger Einstein radii, leading to an overestimation of the line-of-sight mass surface density and thus an overestimation of \(H_0\) \cite{Collett2016}.

In contrast, SNe Ia exhibits smoother light curves, with the evolution period of the SNe Ia light curves typically lasting weeks to a month. This characteristic not only significantly reduces observational expenses but also facilitates the measurement of lens system dynamics, as SNe Ia gradually diminish over several hundred days. This helps reconstruct lens systems and helps to resolve the mass-sheet degeneracy \cite{Barnabe2011, Yildirim2017, Shajib2020, Yildirim2020, Birrer2020}. Furthermore, leveraging the well-established knowledge of regular light curves, gravitationally lensed Type Ia supernovae (glSNe Ia) can be identified without the need to resolve multiple images, thereby reducing the impact of selection biases \cite{Goobar2017}. Third, by using SNe Ia as standard candles, the magnification \(\mu\) caused by gravitational lensing can be directly inferred from their light curves. This offers an alternative method to address the mass-sheet degeneracy \cite{Oguri2003}.  Moreover, time delays in strong lensing provide an opportunity to examine the initial stages of supernova explosions. It is widely acknowledged that the characteristics of SNe in their early phases (usually within 10 rest-frame days) are essential for identifying their progenitors, surroundings, and details of their explosions \cite{Kasen2010,Bulla2020,Li2024}. Additionally, glSNe Ia entails a lower observational time investment than glQSOs.

In contrast to the numerous glQSOs, only nine glSNe have been observed so far, namely PS1-10afx, SN Refsdal, iPTF16geu, SN Requiem, AT 2022riv, C22, SN Zwicky, SN H0pe and SN Encore \cite{Chornock2013,Kelly2015,Goobar2017,Rodney2021,Kelly2022,Chen2022,Goobar2022Zwiky,Pierel2024h0pe,Pierel2024encore}. Among these, PS1-10afx, SN Zwicky, and iPTF16geu occur on the galaxy scale, while the remainder occurs on the scale of groups or clusters. Notably, SN Requiem and SN Encore are found within the same host galaxy. SN H0pe is notable as it is the first SNe Ia with adequately sampled light curves and sufficiently extended time delays to enable an \(H_0\) measurement, leading to an estimation of \(H_0=75.4_{-5.5}^{+8.1} \text{km} \, \text{s}^{-1} \, \text{Mpc}^{-1}\) \cite{Pascale2024}. 

Although, SN Refsdal, identified as a core-collapse SN (CCSN) \cite{Kelly2023a,Kelly2023b,Chen2022}, exhibited extended time delays and well-sampled light curves, enabling \(H_0\) estimation with relatively low uncertainty. This supernova allowed for a determination of \(H_0\) with an uncertainty $\sim 6\%$ within the context of a flat \(\Lambda\) CDM cosmology, yielding values of either \(H_0=64.8^{+4.4}_{-4.3}\) or \(H_0=66.6^{+4.1}_{-3.3} \text{km} \, \text{s}^{-1} \, \text{Mpc}^{-1}\), depending on the assigned lens model weights \cite{Kelly2023a}. Additionally, a value of \(H_0=65.1^{+3.5}_{-3.4} \, \text{km} \, \text{s}^{-1} \, \text{Mpc}^{-1}\) was obtained under more general background cosmological models \cite{Grillo2024}. Nonetheless, as noted previously, the quantity of identified glSNe remains limited, particularly for those with adequately long time delays and well-sampled light curves. The future looks promising with a planned decade-long survey of the entire southern sky, which is expected to result in observations of several hundred lensed SNe \cite{OM10}. And according to Ref. \cite{Dong2024}, the Chinese Survey Space Telescope (CSST) is projected to identify approximately 1009 and 52 strongly lensed supernovae within a decade via its Wide Field Survey (WFS) and Deep Field Survey(DFS), respectively. More glSNe Ia is expected to be observed in the future, expanding the data set and providing better constraints on \(H_0\).

The CSST, a two-meter space telescope, is slated for launch around 2026, sharing an orbit with the Tiangong Space Station. It will be able to dock with the station to facilitate regular or as-needed maintenance, upgrades, and refueling operations \cite{Zhan2021}. Featuring a Cook-type off-axis three-mirror anastigmat design, this telescope will host five instruments: a survey camera, a terahertz receiver, a multi-channel imager, an integral field spectrograph, and a cool-planet imaging coronagraph \cite{Zhan2021}. The core aim of the CSST is to conduct extensive multiband imaging and slitless spectroscopy surveys in parallel with the survey camera. This camera is comprised of 30 detectors, each measuring \(9K \times 9K\) with a pixel size of 0.074 arcsec and offering a cumulative field of view of \(\ 1.1 deg^2\). Of the detectors located at the main focal plane, 18 are designated for multiband imaging covering bands such as $NUV$, $u$, $g$, $r$, $i$, $z$, and $y$, whereas the other 12 are reserved for slitless spectroscopy. The observations, both imaging and spectroscopic, will cover a wavelength span of 255-1000 nm. Imaging observations aim for an \(80\%\) encircled energy radius, \(R_{EE80}\), not exceeding \(0.15''\). The CSST aims to accomplish two distinct programs over ten years: WFS covering 17,500 \(deg^2\), and DFS spanning 400 \(deg^2\). Following the paths of WFS and DFS, each sight-line will be visited on average 2 and 8 times, respectively, in the $u$, $g$, $r$, $i$, and $z$ bands, as well as 4 and 16 times in the $NUV$ and $y$ bands. For these bands, after stacking all exposures, the \(5 \sigma\) limiting magnitudes for point source detection in WFS (DFS) are 25.4 (26.7), 25.4 (26.7), 26.3 (27.5), 26.0 (27.2), 25.9 (27.0), 25.2 (26.4), and 24.4 (25.7) (AB magnitude). The cadence of CSST is anticipated to be approximately 80 days. 
 
The Muztage-Ata One-Ninety-Three-Meter Synergy Telescope (MOST) leaded by Beijing Normal University is a collaborative project by Beijing Normal University (BNU), Xinjiang Astronomical Observatory (XAO), the Nanjing Institute of Astronomical Optics and Technology from the Chinese Academy of Sciences (NIAOT), as well as Xinjiang University (XJU). 
The telescope's photometric capabilities encompass wavelengths ranging from \(3500\) to \(11000\) \AA, simultaneously covering the \(g\), \(r\), and \(i\) bands, and its spectrograph provides three distinct resolutions: \(\delta \lambda/\lambda = 500/2000/7500\). With the aid of a corrective mirror, the telescope's field of view reaches \(20 \times 20 \ arcmin^2\). The \(10\sigma\) limiting magnitude for a 300 seconds exposure in the V-band is 23.79. It utilizes an R-C optical system featuring three focus positions: Cassegrain focus, declination axis focus, and coud\'{e} focus. The telescope's effective aperture is 1.93 meters with a focal ratio of \(f/8\). The CCD pixel size stands at \(13.5 \mu m\). The focal plane has a scale of \(0.183 \ ''\)/pixel and a quantum efficiency of approximately 0.95. Its guiding system ensures tracking accuracy within 0.3 arcseconds for up to 2 hours, while a pointing model correction predicts a pointing accuracy of 5 arcseconds, which can be improved to 1 arcsecond with secondary corrections. Situated in the southwest of the Xinjiang Uygur Autonomous Region of China, at coordinates \(38^{\circ}19'47''\)N, \(74^{\circ}53'48''\)E, and at an elevation of 4526 meters, the Muztagh-Ata site ranks among China's premier astronomical locations. The median seeing value is 0.82 arcseconds \cite{Xu2020b}. During nighttime, the median sky brightness in the V-band is \(21.35 \text{ mag arcsec}^{-2}\), improving to \(21.74 \text{ mag arcsec}^{-2}\) under moonless conditions. Relative humidity averages at \(49\%\) at night and \(39\%\) during the day. The median wind speed at night is \(5.5 \text{ m s}^{-1}\), rising to \(6.5 \text{ m s}^{-1}\) during daylight hours \cite{Xu2020a}. These environmental factors make the telescope particularly well-suited for studies in time-domain astronomy. For example, the potential of MOST for monitoring glQSOs has been investigated in Ref. \cite{Zhu2023}, which concludes that, with a 3-day cadence, MOST can reduce the time delay measurement error to one-third of its current level.

Apart from the sparse data on glSNe Ia, a further challenge in utilizing glSNe Ia is microlensing. It is well recognized that stars within the lensing galaxy have the potential to greatly amplify or diminish the brightness of QSOs or supernovae \cite{Chang1979}. When the same source creates multiple images on the lensing plane, the light takes various routes within the lens system, resulting in a complex magnification pattern over the stellar field. These magnifications vary on microarcsecond scales, generally comparable to the physical size of supernovae and AGNs. Consequently, as a supernova expands on the source plane, each image undergoes unique temporal and wavelength-dependent magnification changes, hindering the alignment of light curves and the extraction of accurate time delays. Moreover, the intrinsic unpredictability of microlensing makes it challenging to incorporate magnification factors of microlensing into the reconstruction of the lens system, as there are inconsistencies between observational data and the model's inadequacies \cite{suyu2024}.

In recent years, multiple studies have concentrated on glSNe Ia with microlensing effects \cite{Goldstein2018,Suyu2020} and have suggested methods to reduce the influence of microlensing on time-delay measurements. One method involves utilizing the color curves of glSNe Ia within the first 15-20 days after the supernova explosion to obtain more precise time delays. The calculation indicates that the uncertainty in relative time delays caused by microlensing is minimal \cite{Goldstein2018,Huber2021}. By fitting the light curves using a Gaussian process in a band that exhibits a sufficiently long achromatic phase of SNe Ia explosion, accurate relative time delays can be obtained.

In this article, we integrate the CSST survey strategy with the follow-up observational capabilities of the MOST telescope to forecast the time delay measurement accuracy for glSNe Ia. Utilizing the well-established W7 supernova explosion model \cite{Nomoto1984}, we perform radiative transfer simulations using the \texttt{SEDONA} code \cite{Kasen2006,Roth2015}. Strong gravitational lensing simulations are generated with \texttt{lfit\_gui} \cite{Shu2016}, while microlensing effects on glSNe Ia light curves are evaluated using the Purple Mountain Observatory's microlensing simulation code \cite{Zheng2022, Zheng2023}. Finally, we measure time delays from the simulated data with \texttt{SNTD} \cite{Pierel2019}. 
The structure of this paper is as follows: Section \ref{sec:2} outlines the strong lensing model, the SNe Ia explosion model, and the lensing population model. Section \ref{sec:3} delves into the microlensing effects on the light curves and color curves. Section \ref{sec:4} details the methodology for fitting light curves and presents the final forecasted results for time delay measurements. Finally, the conclusions are provided in Section \ref{sec:5}. Throughout this study, we adopt a flat \(\Lambda CDM\) cosmological model with parameters \(\Omega_{\Lambda} = 0.74\), \(\Omega_m = 0.26\), and \(H_0 = 72 \, \text{km s}^{-1} \, \text{Mpc}^{-1}\)\cite{Goldstein2018}.

%%%%%%%%%%%%%%%%%%%%%%%%%%%%%%%%%%%%%%%%%%%%%%%%%%%%%%%
\section{Models}\label{sec:2}

In this section, we describe the theoretical framework employed to simulate the global distribution of galaxy-scale glSNe Ia and their associated parameters. We generate velocity dispersion, lensing redshift, lensing axis ratio from the mock catalogs of the CSST's strong lenses \cite{Dong2024}. In addition, we generate shear and convergence from the realistic ellipticity and external shear functions. We randomly assign source positions and solve the lensing equation to determine the imaging location within the multiple-image systems.

%%%%%%%%%%%%%%%%%%%%%%%%%%%%%%%%%%%%%%%%%%%%%%%%%%%%%%%
\subsection{Lensing models}\label{sec:2.1}

Throughout this work, we adopt thin-lens approximation that the deflection effect of the foreground lens only depends on the redshifts of the lens and the source, and the projected mass distribution \( \kappa(\boldsymbol{\theta}) \) in the lens plane perpendicular to the line of sight. Consistent with the notations in Ref. \cite{Narayan1996}, here we denote angular coordinates in the source plane as \(\boldsymbol{\beta}\) and in the image plane as \(\boldsymbol{\theta}\). Then the lens equation describing light deflection is written as:
\begin{equation}\label{eq0}
 \boldsymbol{\beta} = \boldsymbol{\theta} - \boldsymbol{\alpha}(\boldsymbol{\theta})\;,
\end{equation}
where \(\boldsymbol{\alpha}(\boldsymbol{\theta})\) is the reduced deflection angle as a function of \(\boldsymbol{\theta}\). With polar coordinates \(\boldsymbol{\theta}=(\theta\cos\varphi,\theta\sin\varphi)\) We assume that the mass distribution of each elliptical galaxy follows a singular isothermal ellipsoid (SIE), with convergence given by Ref. \cite{Kormann1994}
\begin{equation}\label{eq1}
 \kappa(\theta, \varphi) = \frac{\theta_{\text{Ein}}}{2\theta} \frac{\lambda(e)}{\sqrt{(1-e)^{-1} \cos^{2}\varphi + (1-e) \sin^{2}\varphi}}\;,
\end{equation}
and \(\theta_{\text{Ein}}\) is defined as
\begin{equation}\label{eq2}
 \theta_{\text{Ein}} = 4 \pi \left(\frac{\sigma_v}{c}\right)^2 \frac{D_{\mathrm{ls}}}{D_{\mathrm{s}}}\;,
\end{equation}
where \(D_s\) is the angular diameter distance from the observer to the source plane, and \(D_{ls}\) is the angular diameter distance between the lens plane and the source plane (so that \(D_l\) represents the angular diameter distance from the observer to the lens plane). \(\sigma_v\) denotes the velocity dispersion of the lensing galaxy, \(e\) represents its ellipticity, and \(\lambda(e)\) is the ``dynamic normalization parameter" describing the three-dimensional shape of the lensing galaxy. 

The microlensing magnification map is determined by four key parameters: 1) the convergence field value at the imaging positions; 2) the total shear field value at the imaging positions, which includes both the shear from the strong lens system and the external shear from surrounding structures; 3) the convergence field due to stars at the imaging positions, or the convergence from the luminous (non-smooth) component, \( f_\star = 1 - \kappa_{c}/\kappa \); and 4) the microlens mass function, for which we adopt the Salpeter initial mass function. The relevant parameter in this context is the minimum-to-maximum mass ratio, \( q = m_{\text{min}}/m_{\text{max}} \), within the mass range. Accounting for these effects, the lensing equation becomes

\begin{equation}\label{eq3}
 \boldsymbol{\beta}=\left(\begin{array}{cc}1-\gamma-\kappa_c & 0 \\ 0 & 1+\gamma-\kappa_c\end{array}\right) \boldsymbol{\theta}-\sum_{i=1}^{N_\star} \frac{M_i\left(\boldsymbol{\theta}-\boldsymbol{\theta}_i\right)}{\left(\boldsymbol{\theta}-\boldsymbol{\theta}_i\right)^2}\;,
\end{equation}
where \(\kappa_{c}\) represents the convergence field provided by the smooth matter distribution (or dark matter) at the imaging positions, \(\beta\) is the angular position of the source in the source plane, \(\theta\) is the observed angular position of the source in the lensing plane, \(\theta_i\) is the angular position of the \(i\)-th star, and \(N_\star\) is the number of stars in the local field provided by the local convergence. We used ray-tracing codes from the Purple Mountain Observatory \cite{Dong2024,Zheng2022,Zheng2023} to calculate the microlensing magnification map for each macro images. In this approach, non-smooth convergence fields of stars are treated as point masses randomly distributed within the magnification distribution map. The size of the magnification distribution map is typically described in terms of the Einstein radius \(R_E\) associated with a point mass model

\begin{equation}\label{eq4}
  \bar{R}_E=\sqrt{\frac{4 G \bar{m}}{c^2} \frac{D_{ls} D_s}{D_l}}\;,
\end{equation}

The parameters \( \kappa \) and \( \gamma \) can be directly determined for each imaging positions through the strong lensing simulation process discussed above according to the SIE model and external shear.  \(f_*\) is estimated using the method of Ref. \cite{Dobler2006}, assuming the de Vaucouleurs galaxy profile. In our microlensing simulation, we model stars using the Salpeter initial mass function \cite{1955ApJ...121..161S}, characterized by the mass distribution \(dn/dm \propto m^{-2.35}\). The same initial mass function was employed in our simulation to model the supernova event rate. According to Ref. \cite{Salpeter2005}, we take the Salpeter mass function, with a range from 0.1 to 10 solar masses and an average mass \(\bar{m} = 0.3 M_{\odot}\), which is suitable for old star populations in elliptical galaxies. 

%%%%%%%%%%%%%%%%%%%%%%%%%%%%%%%%%%%%%%%%%%%%%%%%%%%%%%%
\subsection{SNe Ia model}\label{sec:2.2}

Our simulation process for SNe Ia follows that of previous works by \texttt{HOLISMOKES} \cite{Suyu2020, Huber2019}. We used the well-established spherically symmetric W7 model \cite{Nomoto1984} to derive the spectra and light-curve of the SNe Ia. This model represents a one-dimensional explosion model of a white dwarf of carbon-oxygen that closely approaches the Chandrasekhar mass ( \( M_{Ch} \) CO WD). It contains \(0.59M_\odot\) of \(^{56}\)Ni and is known to reproduce crucial observational data for normal SNe Ia, including spectra \cite{Jeffery1995, Nugent1997, Baron2006}. Generally, we assume that the supernova reaches the phase of homologous expansion across its various shells within a few seconds of the explosion. This enables the use of the Monte Carlo radiation transfer code \texttt{SEDONA} \cite{Kasen2006,} to compute the model's spectral time series. 

Under the spherical symmetry, the emitted radiation intensity within two-dimensional shell (\( P_i\)) can be calculated from the escaped photons according to the following equation
\begin{equation}\label{eq9}
  I_{\nu, e,P_i}=\frac{1}{4\pi^2}\frac{1}{(p^2_{i,out}-p^2_{i,in})}\left(\frac{\mathrm{d} E}{\mathrm{d} t \mathrm{d} \nu}\right)_{P_i}\;, 
\end{equation}
where \( p_{i,out}\) and \( p_{i,in}\) represent outer and inner radius of the \( i\)-th shell \( P_i\), and \( \mathrm{d} t\) and \( \mathrm{d} \nu\) denote the rest-frame time and frequency ranges, respectively. 
Hence, we can express the observed supernova intensity \( F_{\lambda,o,t} \) at the observer time $t$ and at given wavelength $\lambda$ as follows
\begin{equation}\label{eq:intensity}
F_{\lambda, o, t} 
= \frac{1}{D_L^2 (1 + z)} \int_0^{2\pi} \mathrm{d}\phi \int_0^{p_{\text{max}}} \mathrm{d}p \, p I_{\lambda, e, p, t}\;,
\end{equation}
where \(\phi \) and \(p \) represent polar coordinates once the supernova has been projected onto a plane. \(I_{\lambda, e,p,t} \) denotes the emitted intensity of light for the specific wavelength \(\lambda \) at time \(t \), and at the distance \(p \). Here, \(D_L \) denotes the luminosity distance to the supernova from the observer, and \(p_{\mathrm{max}} \) indicates the maximum radius. The subscripts \(o\) and \(e\) refer to the light emitted from within the supernova and the light detected outside of it, respectively. In our calculations, we divide the projected sphere into several annuli \(P_i \) with uniform radii. Additional details will be discussed in Section \ref{sec:3.1} and \ref{app:A}.

\end{multicols}
%%%%%%%%%%%%%%%%%%%%%%%%%%%%%%%%%%%%%%%%
\begin{figure}[H]
   \centering
   \includegraphics[width=0.7\textwidth, angle=0]{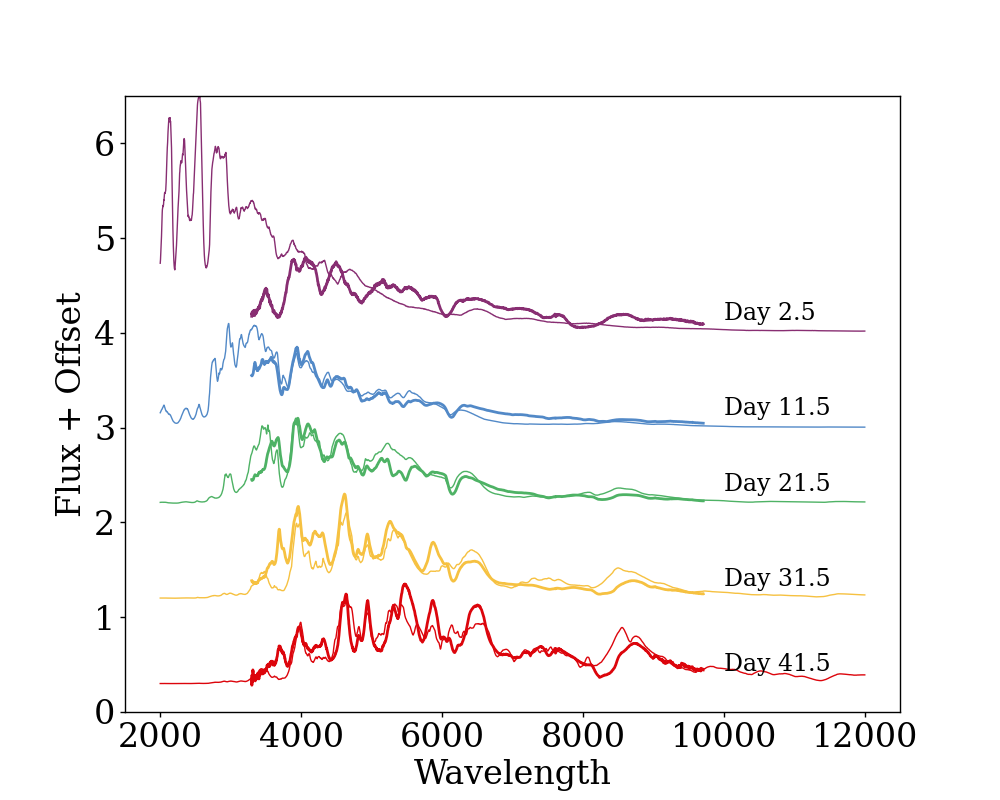}%offsetspectrum.png
   \caption{The normalized spectra for the 2nd, 11th, 21st, 31st, and 41st days following the supernova explosion are derived using the W7 model. The normalization factor is selected based on the flux peak for each corresponding day. For comparison, the spectrum of \texttt{SN 2011fe} is shown as a thick line. To enhance visualization, the spectra are offset from their original values by 4.0, 3.0, 2.2, 1.2, and 0.3, corresponding to the 2nd, 11th, 21st, 31st, and 41st days, respectively. The units for the spectra of the model and 2011fe are \(erg \, s^{-1} \, cm^{-2} \, \text{\AA}^{-1}\).}
   \label{Fig1}
\end{figure}
\begin{multicols}{2}

In our simulation, we calculated the energy and deposition rates of \( \gamma \)-rays produced by the decay of radioactive isotopes \(^{56}\)Ni and \(^{56}\)Co over time, which contribute to the emitted light. We excluded other potential decay elements such as \( ^{48}\)Cr, \( ^{48}\)V, \( ^{48}\)Ti. The radiation heating and cooling rates, along with the temperature structure, were assessed using a Monte Carlo estimator. We utilized \texttt{SEDONA} for iterative calculations to derive the temperatures corresponding to each shell in the initial input model, assuming local thermodynamic equilibrium. Moreover, we incorporated bound-free, free-free, bound-bound, and electron-scattering processes. We also applied the two-line atomic approximation method, presuming all lines to be "pure absorption" lines, which implies that in the two-level atomic form, the ratio of redistribution probability to pure scattering probability across all lines is \( \epsilon_{th} = 1\).

The following section outlines the parameter configurations used in the simulations:

-Model spatial grids: The model's matter distribution is segmented into 108 evenly divided radial zones, attaining a maximum velocity of around \(4.04 \times 10^4 \, \text{km} \, \text{s}^{-1}\). The outer layers of the model are filled with elements \(C\) and \(O\).
  
-Time: The time evolution is tracked across 459 intervals, spanning from day 1 to day 116. The day 0 is selected as the reference point since the SN explosion. Time steps follow a logarithmic increment, given by \( t_2 = t_1(1 + 0.175) \), with a maximum step size of 0.25 days.

-Wavelength: The wavelength range extends from \(300 \, \text{\AA}\) to \(30,000 \, \text{\AA}\), with a resolution of \(10 \, \text{\AA}\). An extensive review confirmed that this resolution is appropriate for the simulations.

-Atomic Line Data: The atomic line data are sourced from the Kurucz CD 23 line table \cite{Kurucz1995}, which contains approximately 500,000 lines. Of these, 300,000 lines are used in the simulations.

-Energy-Momentum Segmentation: The intensity is derived from the output photon packets, with parameter \(p\) segmented into 100 bins, utilizing \(10^9\) photon packets for the simulations. This ensures adequate resolution for the light curves, spectra, and two-dimensional supernova profiles based on velocity.

The resulted normalized spectra for the 2nd, 11th, 21st, 31st, and 41st days after the supernova explosion are depicted with colored thin curves in Fig. \ref{Fig1} (the day of the SN explosion is designated as day 0 for both the model and SN 2011fe). The original unit of the spectra is \(erg \, s^{-1} \, cm^{-2} \, \text{\AA}^{-1}\). The normalization factors are chosen based on the flux peak for each respective day. 

\end{multicols}
%%%%%%%%%%%%%%%%%%%%%%%%%%%%%%%%%%%%%%%%
\begin{figure}[H]
   \centering
   \includegraphics[width=0.7\textwidth, angle=0]{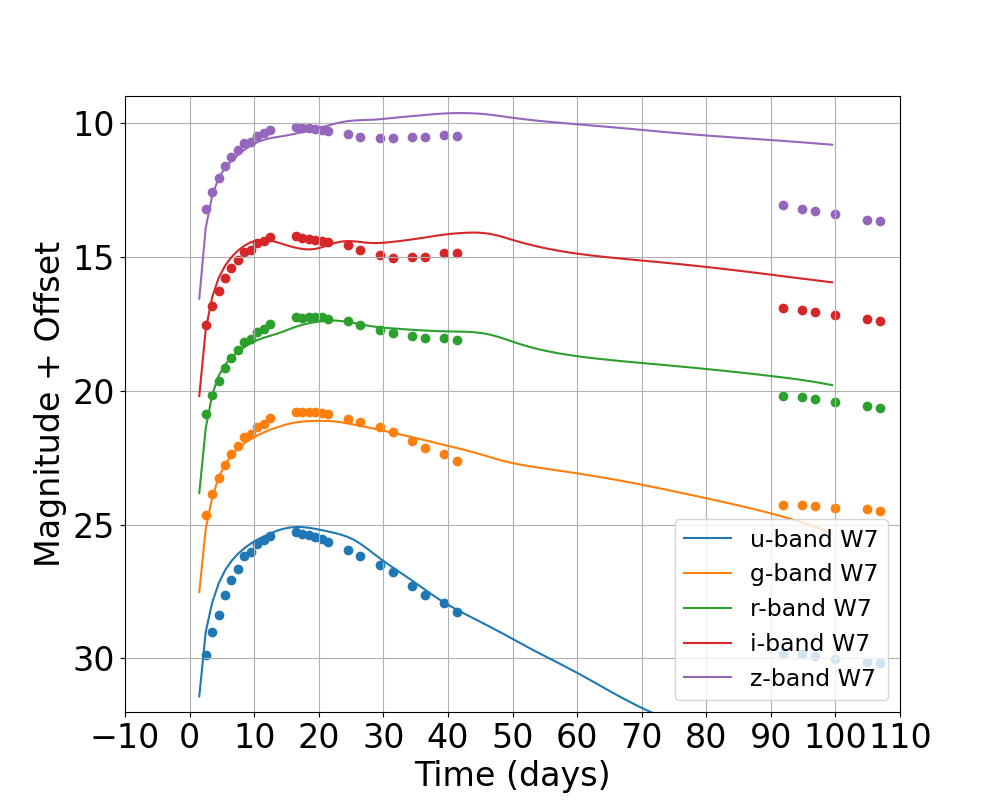}%offsetspectrum.png
   \caption{The apparent magnitudes of the SNe Ia light curves in the Sloan $u$, $g$, $i$, $r$, and $z$ bands are presented. The source redshift is set to that of \texttt{SN 2011fe}, i.e., $z = 0.001208$. The solid curves represent our simulated results, while the points correspond to the observed data for \texttt{SN 2011fe}. The magnitudes are offset relative to their original values by 0, 4, 8, 12, and 16, corresponding to the $z$, $i$, $r$, $g$, and $u$ bands, respectively.}
   \label{Fig2}
\end{figure}
\begin{multicols}{2}

For comparison, the spectrum of \texttt{SN 2011fe}\cite{Pereira2013} is shown as a thick line. To better align the simulated spectra with the observational data, we apply scaling factors of $2.5$, $1.1$, $0.9$, $0.9$, and $0.9$ for the simulated data, and $0.8$, $0.85$, $0.9$, $1.1$, and $1.05$ for the observational data on the corresponding days. Additionally, to improve the visual clarity, the spectra are offset from their original values by 4.0, 3.0, 2.2, 1.2, and 0.3 for the 2nd, 11th, 21st, 31st, and 41st days, respectively.

Utilizing the efficiency of Sloan $u$, $g$, $r$, $i$ and $z$ filters alongside the model spectra, we can generate the light curves shown in Fig. \ref{Fig2}, where the source redshift set to that of \texttt{SN 2011fe} with redshift $z = 0.001208$. The solid curves show our simulated apparent magnitudes, while the data points represent the observed magnitudes for \texttt{SN 2011fe}. For better visibility, we apply a \(\Delta mag= 0, 4, 8, 12\) and \(16\) to the $z$, $i$, $r$, $g$, and $u$ bands, respectively.

%%%%%%%%%%%%%%%%%%%%%%%%%%%%%%%%%%%%%%%%%%%%%%%%%%%%%%%
\subsection{Population model}\label{sec2.3}

Accurately simulating the characteristics of lensing galaxies and the frequency of Type Ia supernovae (SNe Ia) is crucial for predicting the galactic SNe Ia rate. To accomplish this, we used forecasts from Ref. \cite{Dong2024} to estimate the expected number of strong lens systems in the CSST survey. In our model, we represent the lens galaxy as an elliptical galaxy, following the approach outlined in Ref. \cite{Chae2003}. We assume that the ellipticity parameter \( e \) follows a normal distribution within the range [0.0, 0.9], with a mean \( \mu_e = 0.3 \) and a standard deviation \( \sigma_e = 0.16 \). As a first step, we incorporate the velocity dispersion profile typically associated with elliptical galaxies, using an adjusted velocity dispersion function (VDF) that accounts for the unique properties of these systems
\begin{equation}\label{eq6}
  \phi(\sigma,z_l)=\phi_{*}(z_l)\left(\frac{\sigma}{\sigma_{*}(z_l)}\right)^{a} \exp \left[-\left(\frac{\sigma}{\sigma_{*}(z_l)}\right)^{b}\right]\frac{1}{\sigma}\;.
\end{equation}
The parameters are based on Ref. \cite{Dong2024,Geng2021,Yue2022}, with \( [\phi_{*}, \sigma, a, b]=[6.92 \times 10^{-3}(1+z_l)^{-1.18}\mathrm{Mpc}^{-3}, 172.2(1+z_l)^{0.18} \mathrm{~km} \mathrm{~s}^{-1}, -0.15, 2.35]\).
The variation of VDF with redshift is consistent with observed VDFs up to a redshift of \( z = 1.5 \). In this study, the lensing galaxies have redshifts ranging from 0 to 2, with velocity dispersions between \( 100 \, \mathrm{km} \, \mathrm{s}^{-1} \) and \( 450 \, \mathrm{km} \, \mathrm{s}^{-1} \).
For the external shear, we assume that \( \log_{10} \gamma_{\text{ext}} \) follows a normal distribution with a mean of \( \mu = -1.3 \) and a standard deviation of \( \sigma = 0.2 \). This assumption is supported by N-body simulations and semi-analytic models of galaxy formation \cite{Holder2003}.

Having established the properties of the lens galaxies, we now turn to the incorporation of the SNe Ia rate. Unlike the supernova rate employed in \cite{Dong2024}, we adopt and use the following rate model:
\begin{equation}\label{eq7}
 n_{\mathrm{Ia}}=\eta k_{\mathrm{Ia}} \frac{\int_{0}^{t(z \mathrm{sN})} \psi\left[z\left(t-t_{d}\right)\right] f\left(t_{d}\right) d t_{d}}{\int_{0}^{t(z=0)} f\left(t_{d}\right) d t_{d}}~~ \left[\mathrm{yr}^{-1} \mathrm{Mpc}^{-3}\right]\;,
\end{equation}
where \( \psi(z) \) represents the star formation history, \( f(t_d) \) denotes the delay-time distribution \cite{Maoz2012}, \( \eta \) is the fraction of progenitors that successfully explode as SNe Ia, and \( k_{\mathrm{Ia}} \) indicates the number of SNe Ia progenitors per unit of stellar mass formed. The star formation history used in this study is based on the model from \cite{Madau2014}, which was derived by fitting observational data from the ultraviolet (UV) and infrared (IR) spectra, covering sources with a redshift \( z_s \sim 3.5 \). The parameter \( \eta \) is set to a canonical value of 0.04 \cite{Hopkins2006, Maoz2012}. The value of \( k_{\mathrm{Ia}} = 0.021 \, \mathrm{M}_{\odot}^{-1} \) is derived using the stellar initial mass function (IMF) from \cite{Salpeter1955}, considering a mass range from \( M_{\min} = 0.1 \, \mathrm{M}_{\odot} \) to \( M_{\max} = 125 \, \mathrm{M}_{\odot} \). The assumed progenitor masses for SNe Ia are between \( 3 \, \mathrm{M}_{\odot} \) and \( 8 \, \mathrm{M}_{\odot} \).

Simulations suggest that approximately 1,105 normal Type Ia supernovae are expected to occur annually per square degree \cite{Dong2024}. Our methodology is as follows: For each lensing galaxy at redshift \( z_l \), we randomly select 1,105 supernovae from the background supernova sample, assuming they lie within a light cone of one square degree centered on the chosen lensing galaxy. These supernovae are randomly distributed within this area. For any supernova at a higher redshift than \( z_l \) and located within a projected distance of less than \( 8 \theta_{\text{Ein,max}}^2 \) from the lensing galaxy, we apply the lens equation using the \texttt{lfit\_gui} code, as outlined in \cite{Shu2016}. If the supernova is strongly lensed, meaning it generates multiple images, the system is included in our simulated catalog. The supernova's characteristics, along with those of the multiple images and the corresponding lensing galaxy, are then recorded in the catalog. Our catalogs cover 4000 square degrees. 

We selected the observable events based on the following criteria: 1. Early Detection: The glSNe Ia are detected early, within 1 to 3 days of the supernova explosion, providing high-quality time-series data for the first captured images; 2. Signal-to-Noise Ratio: The MOST observations offer an adequate signal-to-noise ratio, with the V-band peak magnitude of the first arrived image is not fainter than 23; 3. Resolution: The MOST site provides sufficient seeing conditions to resolve the images, with a minimum image separation of \( 0.82" \).
As a result, we identified 2 quadruple-image systems and 14 double-image systems in the 4000 square degrees of the sky observable by MOST each year. This is an initial prediction that omits actual weather, observing strategies, and other potential influencing factors.

For typical redshifts of the source and lens (\( z_s = 1.2\), \( z_l = 0.6\)), the size of SNe Ia five days after the explosion is approximately \( R_{\rm 5days} \approx 7.9 \times 10^{-5} \) pc. At the same redshifts, the Einstein radius for the average mass \( \bar{m} \) is about \( R_E \approx 5.3 \times 10^{-4} \) pc, roughly six times larger than the SNe Ia size. This suggests that the microlensing-induced light variation is significant enough to be detectable. Insufficient resolution in the microlensing magnification map can lead to distorted fluctuations in the color curves during the early phases of glSNe Ia explosions. To address this, we divided the expected supernova intensity into 100 annuli and chose a magnification distribution map with dimensions of \( 20\bar{R}_E \times 20\bar{R}_E \), setting the resolution at \( 4\times10^8\) pixels. This ensures an adequate level of resolution for the magnification distribution map.

%%%%%%%%%%%%%%%%%%%%%%%%%%%%%%%%%%%%%%%%%%%%%%%%%%%%%%%
\section{Light curve and color curve simulations}\label{sec:3}

\begin{figure}[H]
   \centering
   \includegraphics[width=0.9\textwidth, angle=0]{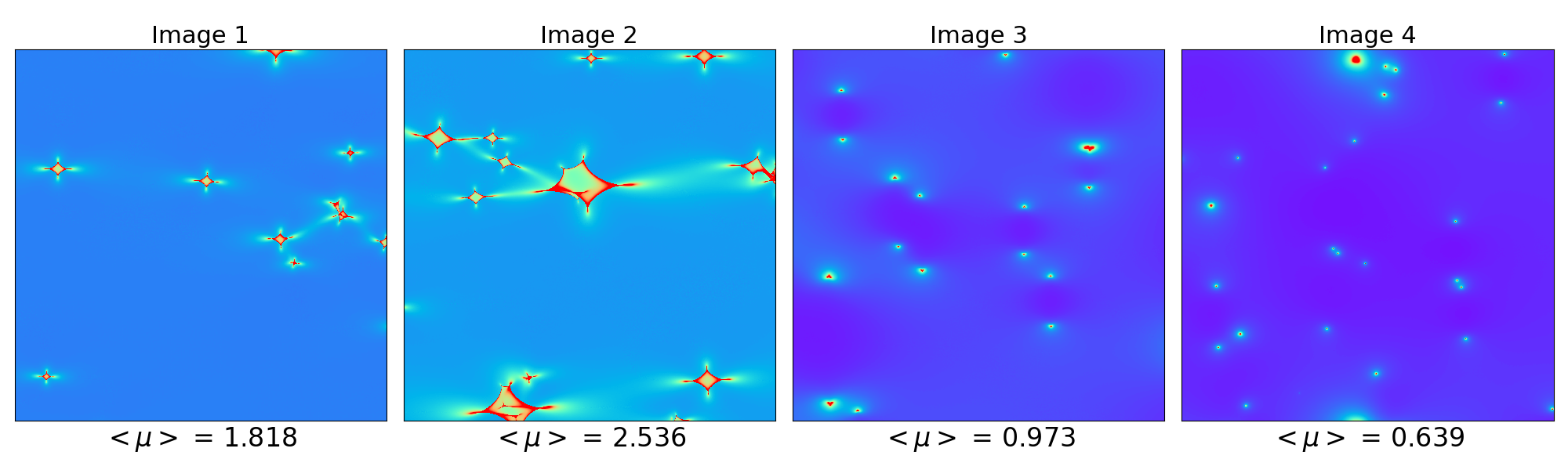}
   \caption{These panels demonstrates the microlensing magnification maps of quadruple-image systems. \( \langle\mu\rangle\) is the mean magnification of each map.}
   \label{fig:4map}
\end{figure}

In this section, we describe the simulation methodology for light curves and color curves, specifically the incorporation of the microlensing effect into glSNe Ia. After fixing the strong lens system parameters, we generate images of the microlensing magnification map and SNe Ia spectra captured under the influence of microlensing effects. In the following simulations, we select one quadruple-image system and one double-image system for demonstration. For quadruple(double)-image systems, the velocity dispersion $\sigma=272.305\;(236.683)\;\text{km/s}$ , ellipticity $e=0.398\;(0.373)$, external shear $\gamma=0.101\;(0.055)$, lens redshift $z_l=0.256\;(0.144)$, and source redshift $z_s=0.495\;(0.381)$, respectively. For each system, we compute the microlensing magnification at multiple positions to capture the stochastic nature of the realizations. The parameters for the strong lensing at different image positions are provided in Table \ref{tab:SL}. For example, the microlensing magnification maps for the quadruple-image system are shown in Figure \ref{fig:4map}.

%%%%%%%%%%%%%%%%%%%%%%%%%%%%%%%%%%%%%%%%%%%%%%%%%%%%%%%
\begin{table}[H]
\centering
\footnotesize
\begin{threeparttable}\caption{$\kappa$: local convergence, $f_*$: stellar fraction. 
$\gamma$: local shear and $\langle \mu \rangle$: macro-magnification. The first and second rows are for double-image system and the third to sixth rows are for the quadruple-image system. $No$: index of images. These numeral sets are employed throughout this research to differentiate various images.}\label{tab:SL}
\doublerulesep 0.1pt \tabcolsep 12pt %space between two columns. 
\begin{tabular}{ccccccc}
\toprule
 \hline
 No & $\mu$ & $\kappa$ & $\gamma$ & $f_*$ \\
 \hline
1 & 2.599 & 0.300 & 0.324 & 0.047  \\
2 & -2.365 & 0.673 & 0.727 & 0.146  \\
 \hline
1 & 1.866 & 0.199 & 0.241 & 0.157 \\
2 & 2.477 & 0.231 & 0.323 & 0.155  \\
3 & -1.097 & 0.957 & 0.922 & 0.089  \\
4 & -0.767 & 1.143 & 1.057 & 0.123  \\
 \hline
\bottomrule
\end{tabular}
\end{threeparttable}
\end{table}

%%%%%%%%%%%%%%%%%%%%%%%%%%%%%%%%%%%%%%%%%%%%%%%%%%%%%%%
\subsection{Microlensing}\label{sec:3.1}

We follow Ref. \cite{Goldstein2018} to cooperate the microlensing magnification. The observed flux \( F_{\lambda,\mathrm{o}}\) is obtained by convolving the emitted intensity \( I_{\nu, \mathrm{e},P_i}\) with the corresponding microlensing magnification map
\begin{equation}\label{eq10}
  \mu F_{\lambda, \mathrm{o}}=\frac{2\pi(1+z)}{D_L^2} \int_0^{P_{\mathrm{S}}} \mathrm{d} P~\mu_P P I_{\lambda, \mathrm{e},P}\;,
\end{equation}
where \(\mu_{P}\) represents the mean magnification within the projected shell \(P_i\) on the microlensing magnification map. \(P_S\) denotes the outermost radius of the SNe Ia photon sphere.
Due to the inherent uncertainty in the spatial positioning of the microlenses, we simulate the strong lensing system multiple times by placing the supernova at different locations within the same microlensing magnification map. Here, we only model the uniform expansion of the supernova with the magnification (i.e., the change in the size of the supernova projection on the magnification over time) rather than the relative motion between the supernova and the lensing galaxy field.

\end{multicols}

%%%%%%%%%%%%%%%%%%%%%%%%%%%%%%%%%%%%%%%%

\begin{multicols}{2}

\end{multicols}

%%%%%%%%%%%%%%%%%%%%%%%%%%%%%%%%%%%%%%%%
\begin{figure}[H]
   \centering
   \includegraphics[width=0.7\textwidth, angle=0]{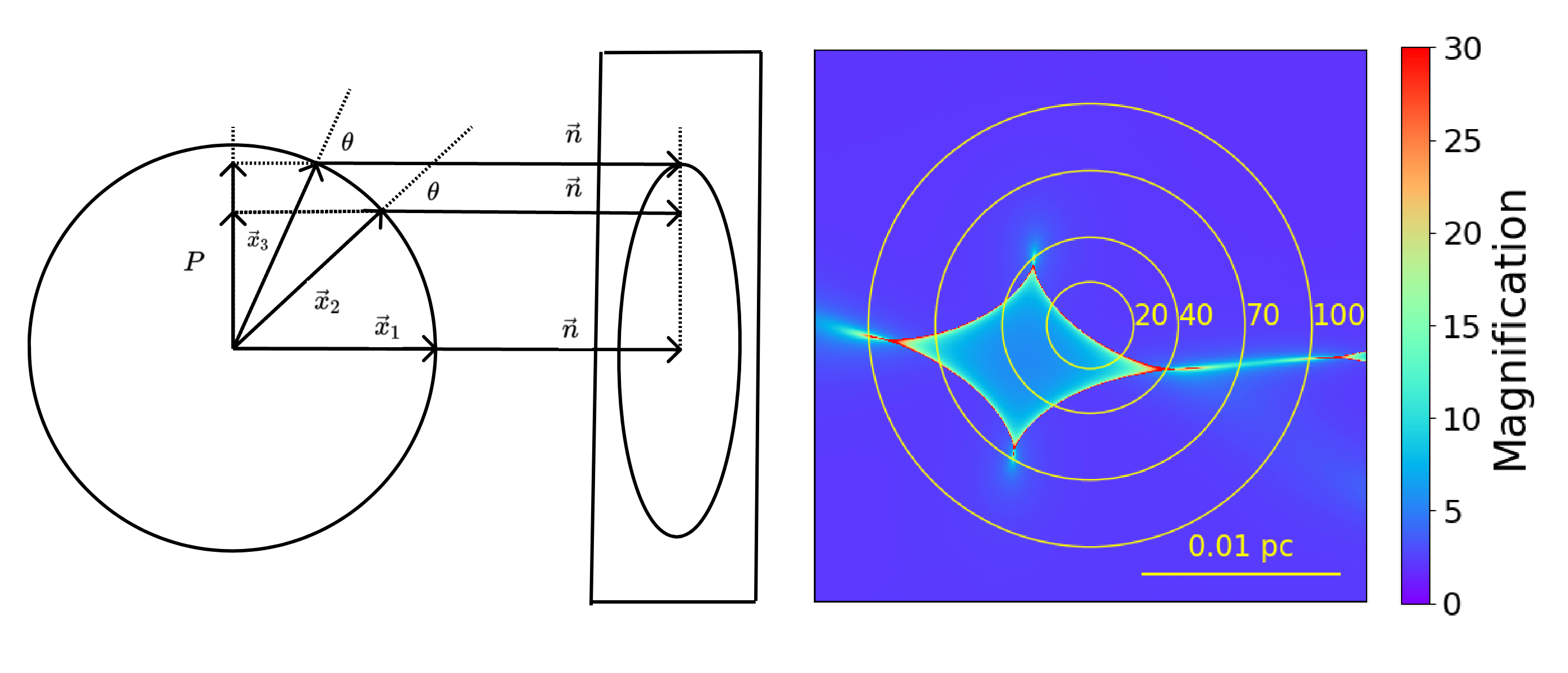}
   \caption{The left panel demonstrates the method of retrieving information about escaped particles using the \texttt{SEDONA} code, producing a two-dimensional intensity map by projecting the supernova. For further details, see the \ref{app:A}. A schematic on the right panel illustrates the supernova's projected maximum radius at different days alongside the microlensing magnification map. Each yellow circle is labeled with a numeral denoting the days elapsed since the explosion. The scale in the lower right corner is for the scenario where \( z_l = 0.256\) and \( z_s = 0.495\). }
   \label{fig:snillu}
\end{figure}

\begin{multicols}{2}

This is because the supernova's atmosphere expands and moves relative to the lensing galaxy, with the expansion velocity of the supernova's atmosphere (\(\sim 10^4 \, \text{km/s}\)) being much higher than the characteristic relative velocity between the supernova and the lensing galaxy (\(\sim 10^2 \, \text{km/s}\)). As a result, we focus solely on simulating the effect of the atmosphere's expansion. To better illustrate the method used to produce the light curve realization, we present Figure \ref{fig:snillu}. The left panel of Figure \ref{fig:snillu} demonstrates the projection effect of photons emitted from a 3D spherical photon sphere. For further details, refer to \ref{app:A}. The right panel of Figure \ref{fig:snillu} provides a schematic representation of microlensing caustics crossing through different projected shells.

\begin{figure}[H]
   \centering
   \includegraphics[width=0.7\textwidth, angle=0]{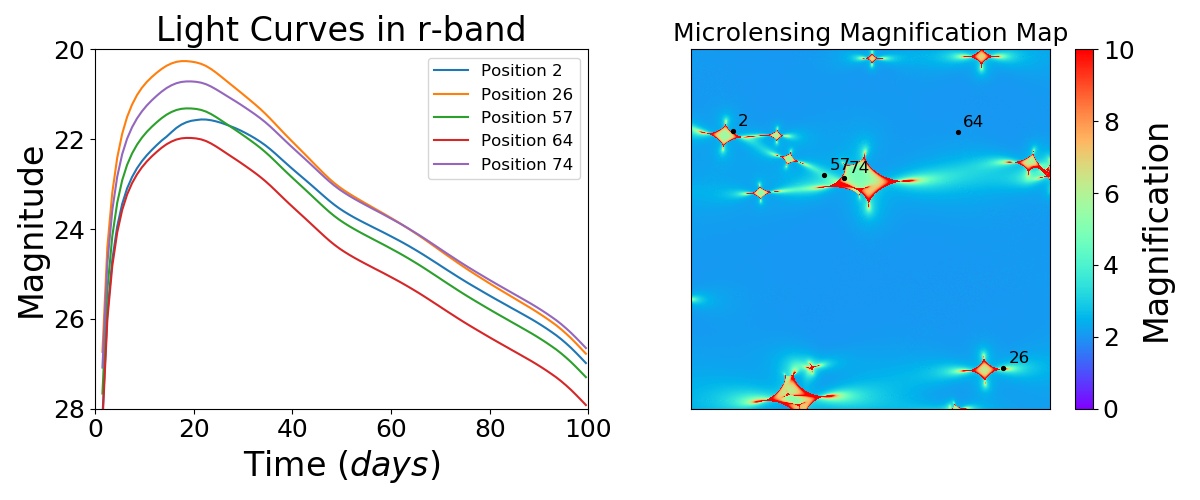}
   \caption{The left panel shows the simulated \(r\)-band light curves for different supernova positions within the microlensing magnification map. The right panel provides a schematic representation of these positions on the map, with the numbers next to the points indicating the indices of the corresponding light curve realizations. The map is generated using parameters from the second imaging of the quadruple-image lens system.}
   \label{fig:lc_micro}
\end{figure}

The left panel of Figure \ref{fig:lc_micro} presents the $r$-band light curves for supernovae located at different positions within the microlensing magnification map. The right panel shows a schematic of the corresponding locations of these supernovae within the map. When a supernova aligns with the microlensing caustics, its peak brightness can be amplified by up to a factor of approximately 5. The intrinsic magnitudes of all the light curves are identical; the observed differences reflect the effects of microlensing magnification.

\begin{figure}[H]
\centering
\includegraphics[width=0.8\textwidth, angle=0]{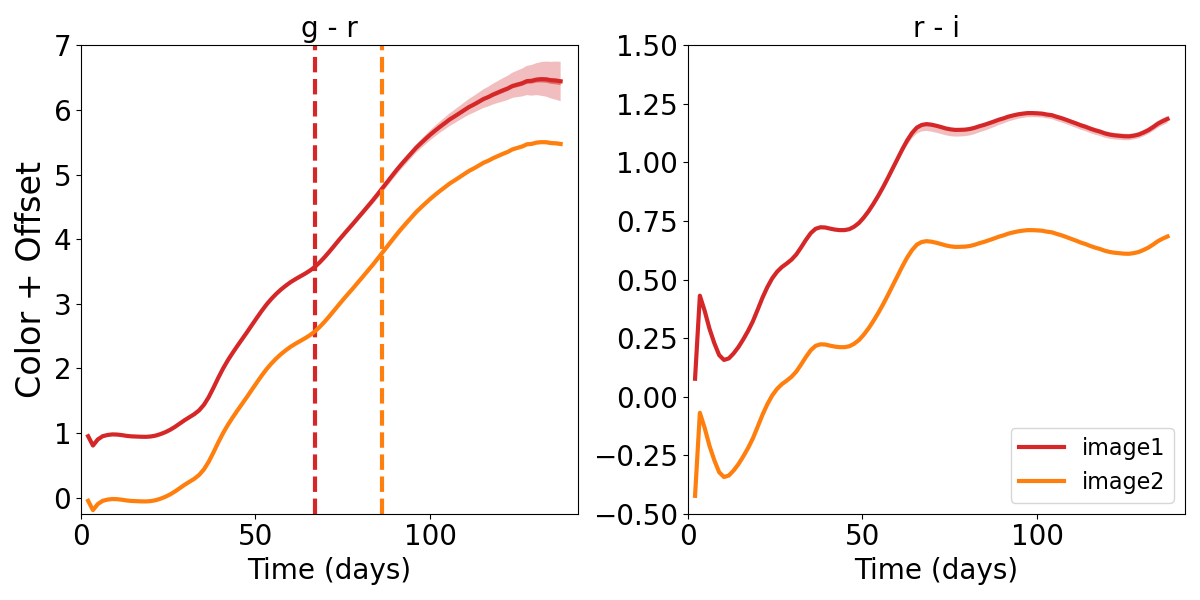}
\caption{The color curves for a double-image system are shown, with an additional offset included in the curves. The \(1\sigma\) and \(2\sigma\) confidence intervals, resulting from the microlensing effect, are also represented for 400 color curves. Vertical dashed lines indicate the onset of the chromatic phase. The left panel displays the $g-r$ color curves, while the right panel shows the $r-i$ color curves.}
\label{fig:2ico}
\end{figure}

\begin{figure}[H]
\centering
\includegraphics[width=0.8\textwidth, angle=0]{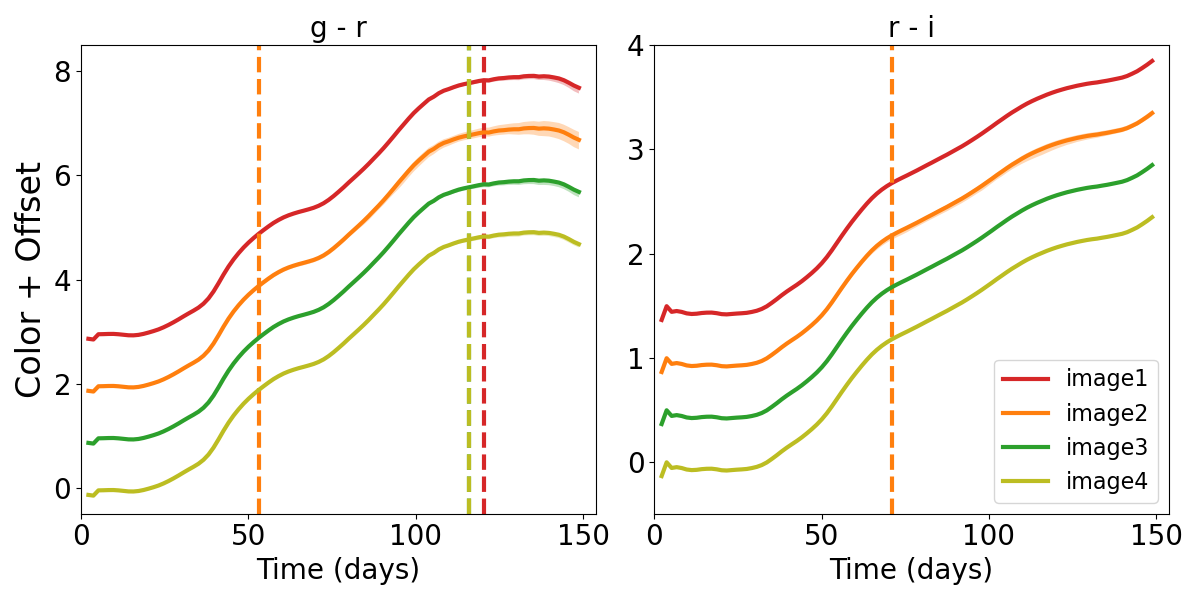}
\caption{Similar to Figure \ref{fig:2ico}, this figure shows the color curves for the quadruple-image system.}
\label{fig:4ico}
\end{figure}

Figures \ref{fig:2ico} and \ref{fig:4ico} present the color curves for the double-image and quadruple-image systems, respectively. The left panels illustrate the \( g - r\), while the right panels display the \( r - i\). The vertical lines follow the definition provided by Ref. \cite{Huber2021} to mark the boundary between the achromatic and chromatic phases of SNe Ia explosion.  The shaded region represents the scatter induced by the microlenses, which is calculated via 400 random SNe Ia positions in magnification map. The microlensing-induced scatter is noticeably more pronounced in the \( g-r \) compared to the \( r-i \). This is because, around 7000 K, iron undergoes a transition from Fe-III to Fe-II, as described by Ref. \cite{Kasen2006, Kasen2007}. 

\end{multicols}

%%%%%%%%%%%%%%%%%%%%%%%%%%%%%%%%%%%%%%%%

\begin{multicols}{2}

%%%%%%%%%%%%%%%%%%%%%%%%%%%%%%%%%%%%%%%%%%%%%%%%%%%%%%%%%%%%%%%%%%%%%%%

\end{multicols}
%%%%%%%%%%%%%%%%%%%%%%%%%%%%%%%%%%%%%%%%
\begin{multicols}{2}

\end{multicols}
%%%%%%%%%%%%%%%%%%%%%%%%%%%%%%%%%%%%%%%%

\begin{multicols}{2}

This transition notably increases opacity to blue and ultraviolet radiation, a phenomenon known as line blanketing. This increased opacity allows redder emission from deeper layers within the supernova to become visible, while emission in the blue and ultraviolet spectra is pushed to outer radii. 

Concurrently, a fluorescent shell of recombining iron forms near the onset of the secondary maximum, generating a peak in redder wavelengths in the supernova's specific intensity profile. Together, line blanketing and the fluorescent shell create a spatial variation in the specific intensity ratio across the supernova.

\begin{figure}[H]
   \centering
   \includegraphics[width=0.7\textwidth, angle=0]{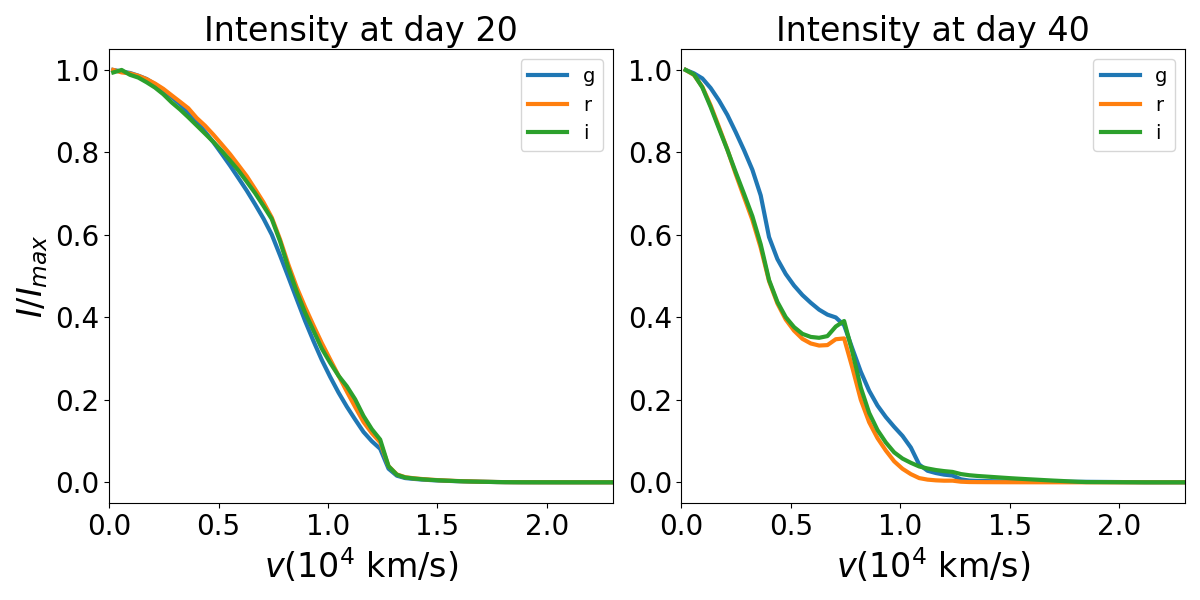}
   \caption{The specific intensity profile \( I_{\lambda(v)}\) for W7 model at 20 days (left panel) and 40 days (right panel) after explosion across the bands of MOST's filters.}
   \label{intensity}
\end{figure}

Figure \ref{intensity} shows the specific intensity curves \( I_{\lambda(v)}\) of the W7 model at 20 days and 40 days after explosion, where \( v\) is the velocity of the shell, equivalent to the radial variable. The graph on the left shows that near the peak (20 days after the explosion), the ratio of light intensity in different bands remains roughly constant at all radial positions. Therefore, ignoring the irrelevant factor of the proportionality coefficient, the supernova near the peak has basically the same intensity curves. Therefore, convolving any magnification pattern \( \mu(P,\phi)\) will not cause significant fluctuations in its color curve, which aligns closely with the simulations presented by Ref. \cite{Goldstein2018}.

\end{multicols}
%%%%%%%%%%%%%%%%%%%%%%%%%%%%%%%%%%%%%%%%

\begin{multicols}{2}

%%%%%%%%%%%%%%%%%%%%%%%%%%%%%%%%%%%%%%%%
\subsection{Photometry calculation}\label{sec:3.2}

\begin{figure}[H]
   \centering
   \includegraphics[width=0.45\textwidth, angle=0]{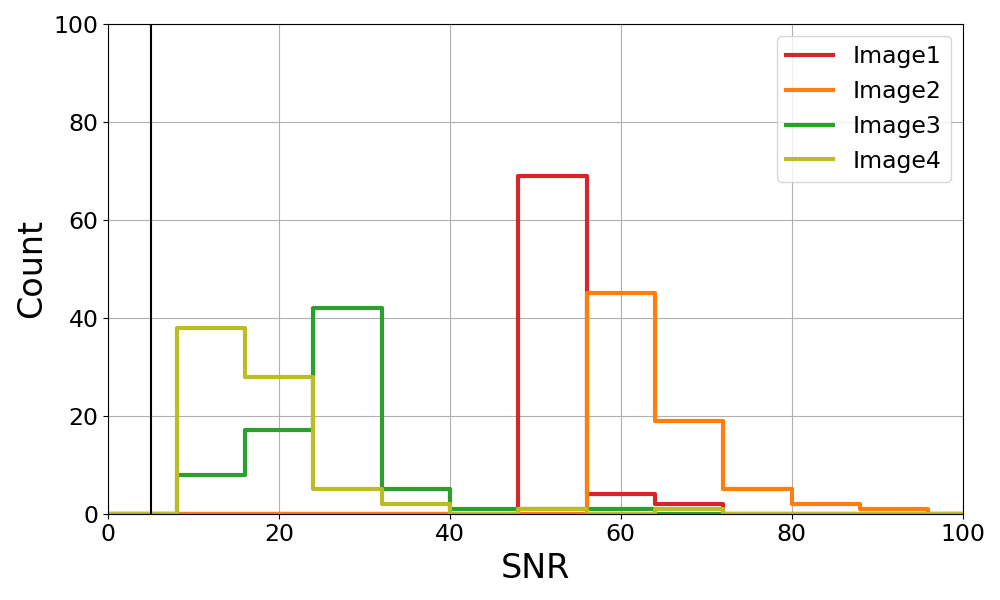}
   \caption{The distribution of the image signal-to-noise ratio (SNR) for 100 simulated quadruple-image systems is shown. The black vertical line indicates the SNR threshold of 5.}
   \label{fig_SNR}
\end{figure}

In this section, we perform the photometric measurements. Typically, the faintest image reaches a peak magnitude of approximately 22 mag. To ensure a well-sampled light curve, we set the exposure time for MOST to 45 minutes. To minimize the effects of varying observing conditions, we divide the total exposure time into 9 slots, each lasting 300 seconds. And we set the cadence as 2 days. Our selection criterion ensures that even the least magnified images yield light curves with a minimum of 5 days of observations, while maintaining a signal-to-noise ratio (SNR) greater than \(5\sigma\). We set the full width at half maximum (FWHM) of the seeing to \(0.82''\), and the median nighttime sky brightness in the \(V\) band is \(21.35\) mag arcsec\(^{-2}\), as reported in previous work \cite{Xu2020b}. This value is then converted from the \(V\) band to the \(r\) band. The SNR is calculated according to
\begin{eqnarray}
\label{eqsnr}
     \mathrm{SNR} 
     %&=& \frac{N_{\text {star }}}{\sqrt{N_{\text {star }}+n_{\text {pix }}\left(N_{\text {sky }}+R_{\text {readout }}^2\right)}} \;,\nonumber\\
     &=&\frac{n_{\text {star } t_{\text {obs }}}}{\sqrt{n_{\text {star }} t_{\text {obs }}+n_{\text {pix }}\left(n_{\text {sky }} t_{\text {obs }}+R_{\text {readout }}^2\right)}}\;,
\end{eqnarray}
where \( n_{\rm pix} \) represents the number of pixels covered by the image, primarily determined by the seeing conditions. \( R_{\rm readout} \) denotes the readout noise, set to \( 5 \, e^- \, {\rm pixel}^{-1} \). The SNR statistics is summarized in Figure \ref{fig_SNR}. It can be observed that the brightest image achieves an SNR of approximately 50, while the faintest image has an SNR of at least 7.

%%%%%%%%%%%%%%%%%%%%%%%%%%%%%%%%%%%%%%%%

%%%%%%%%%%%%%%%%%%%%%%%%%%%%%%%%%%%%%%%%%%%%%%%%%%%%%%%
\section{Light curve fitting}\label{sec:4}

Using the \texttt{SNTD} code \cite{2021ApJ...908..190P,2019ApJ...876..107P}, we applied the \texttt{SALT2} model \cite{Guy2007} to fit each group of simulated light curves. The model served as a template for parameterizing the SNe Ia spectra, with the flux of the template represented by:
\begin{equation}\label{eq12}
F_\lambda(\lambda, t)=x_0\left[M_0(\lambda, t)+x_1 M_1(\lambda, t)\right] \exp [c C L(\lambda)]\;.
\end{equation}
The template includes four parameters: \(t_0\), \(x_0\), \(x_1\), and \(c\), which specify a reference time, a general SED normalization, a supernova "stretch", and a color-law coefficient, respectively. Here, \(M_0\) and \(M_1\) are eigenspectra obtained from a training set of observed SNe Ia spectra, and \(CL(\lambda)\) represents the mean color-correction law of the sample. For the selected sample, the apparent magnitude peaks in the \( r \)-band; therefore, we use the \( r \)-band data for the light curve fitting. For each fitting procedure, we approximate a time delay of
\begin{equation}\label{eqtd}
\Delta t_{i-r} = t_i - t_r\;, 
\end{equation}

\end{multicols}

%%%%%%%%%%%%%%%%%%%%%%%%%%%%%%%%%%%%%%%%%%%%%%%%%%%%%%%
\begin{figure}[H]
   \centering
   \includegraphics[width=0.7\textwidth, angle=0]{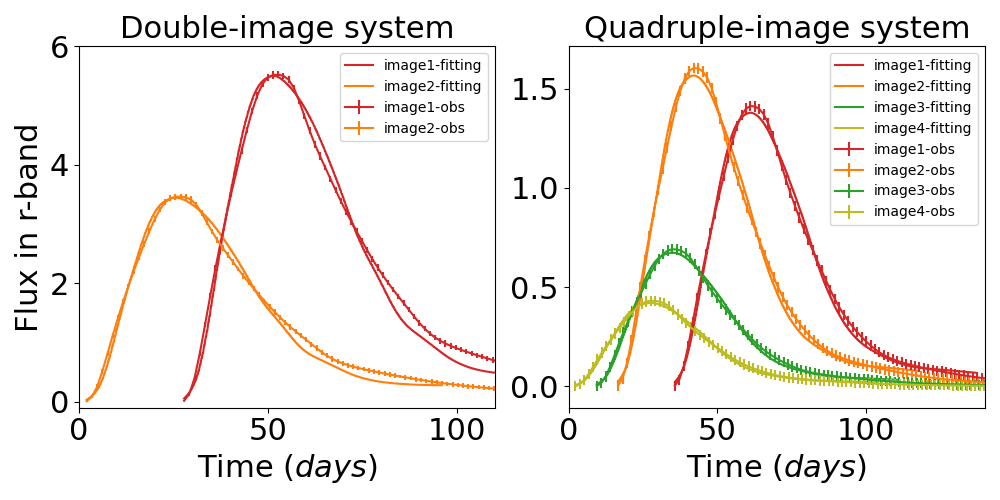}
   \caption{This figure shows an example of a simulated $r$-band light curve alongside its corresponding fit. The fit, generated by the \texttt{SNTD} code, is represented by the solid curve, while the crosses with error bars indicate the simulated observational data. The left and right sub-panels show the results of the double-image and quadruple-image system, respectively.}
   \label{fig:sntd}
\end{figure}

\begin{multicols}{2}

where \( t_r \) denotes the reference time when the signal arrives, and \( t_i \) refers to the time of the \( i \)-th image, respectively. When the photometry SNR is sufficiently high, one should opt for the image least likely to be impacted by microlensing. The choice of reference image can be guided by the scale of color curve fluctuations discussed in Section \ref{sec:3.1} and the timing of chromatic phase occurrences. However, when evaluating these fluctuations, both the apparent magnitude and the lensing magnification must be considered. Generally, higher magnifications increase the possibility of microlensing effects, although this also results in a higher SNR. Assuming SIE model, as indicated in Figure \ref{fig:4ico}, under ideal photometric conditions, either image 3 or image 4 should be chosen as the reference. 

Figure \ref{fig:sntd} illustrates an example of the simulated light curve alongside its fitted counterpart. The \texttt{SNTD} code's best fitting outcome is represented by the solid curve, while the cross with error bars show the simulated observation inputs. Table \ref{tab:fitresult} shows the time delay fitting results of the example glSNe Ia. 

%%%%%%%%%%%%%%%%%%%%%%%%%%%%%%%%%%%%%%%%%%%%%%%%%%%%%%%
\begin{table}[H]
\centering
\footnotesize
\begin{threeparttable}\caption{The time delay fitting results of the example glSNe Ia. \(\epsilon\) denotes middle values  of the biases, while the error indicates median of the errors across different image pairs. \(\Delta t_{true}\) represents the true relative time delay.}\label{tab:fitresult}
\doublerulesep 0.1pt \tabcolsep 12pt %space between two columns.
\begin{tabular}{cccc}
\toprule
 \hline
 No & $\epsilon$~~[days] &  error~~[days] & $\Delta t_{true}$ \\
 \hline
1-2 & -0.028 & $\pm 0.143$ & -25.753\\
 \hline
1-3 & -0.011 & $\pm0.215$ & 26.259\\
2-3 & 0.041 & $\pm0.206$ & 7.188\\
4-3 & 0.012 & $\pm0.357$ & -7.232\\
 \hline
\bottomrule
\end{tabular}
\end{threeparttable}
\end{table}

By combining Figure \ref{fig:4map} and Table \ref{tab:SL}, one can see that the errors in light curve fittings are typically around a few hours, with biases generally being under one hour. This finding suggests that the time delay bias caused by microlensing will not significantly impact the systematic accuracy of cosmography utilizing glSNe Ia.

%%%%%%%%%%%%%%%%%%%%%%%%%%%%%%%%%%%%%%%%%%%%%%%%%%%%%%%
\section{Conclusion}\label{sec:5}

Strong lensing time delays, including those from glQSOs and glSNe Ia, serve as a powerful tool for measuring cosmological distances, the size of black hole accretion disks, exploring the explosion mechanism of SNe Ia and more. The Muztagh-Ata 1.93-meter Synergy Telescope (MOST) is an outstanding instrument for conducting such measurements. Compared to glQSOs, glSNe Ia offer several advantages: more regular light curves, reduced contamination from microlensing, superior photometric observation conditions, etc.
In this paper, we forecast the time delay measurement accuracy that can be achieved by MOST for glSNe Ia. 
We simulate the SNe Ia explosion, strong lensing population model and derive the detection rate of glSNe Ia per year according to MOST observable condition.  On average, our findings indicate the potential to detect 2 quadruple-image systems and 14 double-image systems annually.

In addition to the macro magnification caused by strong lensing, the microlensing effects induced by the stellar field within the lens galaxy can also alter the light curve. We simulate the microlensing magnification map based on a realistic scenario and generate microlensed light curves and color curves hundreds of times to account for uncertainties. Finally, we select one quadruple-image system and double-image system as an example to demonstrate the capability of MOST for measuring the glSNe Ia time delay. We adopt the standard \texttt{SALT2} model for the SNe Ia light curve template and use \texttt{SNTD} code for measuring the time delay. Since the typical glSNe Ia is faint, with the peak magnitude of approximately 22 mag. To ensure a well-sampled light curve, we set the exposure time for MOST to 45 minutes.
From the examples, we observe that the initiation of the chromatic phase for certain images is either relatively delayed or entirely absent. In both two sampled systems, the earliest onset of the chromatic phase occurs after day 50. Light curves of some images are not subject to strong fluctuation, indicating that the time delay derived from the light curves fitting is sufficient. Under these conditions, we find that the time delay errors are typically around a few hours, with biases generally being under one hour. 
This study's methodology for evaluating microlensing effects on time delays can be applied to CCSNe but is dependent on the specific supernova progenitor model. According to Ref. \cite{OM10,Dong2024}, hundreds of glSNe are anticipated annually, with approximately three-quarters being CCSNe. Given that CCSNe are brighter than SNe Ia, these systems hold promise for MOST. Furthermore,MOST can conduct follow-up observations of glSNe detected in the Zwicky Transient Facility (ZTF) survey\cite{ZTF2019} or by employing observations from the Wide Field Survey Telescope (WFST)\cite{WFST2023}, extending beyond the CSST survey program.

In summary, MOST is an exceptional instrument for monitoring glSNe Ia(CC) and has the potential to provide crucial evidence in resolving the Hubble tension.

%\section{References}\label{sec:8}

%%%%%%%%%%%%%%%%%%%%%%%%%%%%%%%%%%%%%%%%%%%%%%%%%%%%%%%
%%% Acknowledgements. 
%%%%%%%%%%%%%%%%%%%%%%%%%%%%%%%%%%%%%%%%%%%%%%%%%%%%%%%
\section*{Acknowledgments} 
This work was supported by the National Natural Science Foundation of China (Grant No. 12333001, 61234003, 61434004, 61504141), by CAS Interdisciplinary Project (Grant No. KJZD-EW-L11-04) and by China Manned Space Program through its Space Application System. BH acknowledges the hospitality of the International Centre of Supernovae (ICESUN), Yunnan Key Laboratory at Yunnan Observatories Chinese Academy of Sciences. Z.H. acknowledges support from the China Postdoctoral Science Foundation under Grant Number GZC20232990 and the National Natural Science Foundation of China (Grant No. 12403104).

%%%%%%%%%%%%%%%%%%%%%%%%%%%%%%%%%%%%%%%%%%%%%%%%%%%%%%%
%%% Conflict of interest. 
%%%%%%%%%%%%%%%%%%%%%%%%%%%%%%%%%%%%%%%%%%%%%%%%%%%%%%%
\section*{InterestConflict}
The authors declare that they have no conflict of interest.

\printbibliography

%%%%%%%%%%%%%%%%%%%%%%%%%%%%%%%%%%%%%%%%%%%%%%%%%%%%%%%
%%% Appendix sections. ??????, ????
%%%%%%%%%%%%%%%%%%%%%%%%%%%%%%%%%%%%%%%%%%%%%%%%%%%%%%%
\begin{appendix}
%\section{Name}

%\end{appendix}

%\begin{appendices}
%\section{Appendix}
%\end{appendices}
%\appendix

%\appendix

\renewcommand{\thesection}{Appendix}

\section{}\label{app:A}

In this appendix, we provide a detailed description of the parameters and procedures used to simulate radiation transport, compute the evolving SNe Ia SED (as discussed in Section \ref{sec:2.2}), and outline the method presented in Section \ref{sec:3.1} for combining the intensity with a microlensing magnification map. We utilized the open-source, 3D Monte Carlo radiation transport code *SEDONA* \cite{Kasen2006sedona}, which is capable of modeling the time evolution of supernovae. This code calculates the light curves and spectra of supernova explosion models by statistically tracking the escape of particles (i.e., photons). As mentioned in Section \ref{sec:2.2}, we employed the well-known W7 model for our simulations. By using this specific model, we can define a homologous expansion structure for the supernova ejecta, and *SEDONA* will simulate the average spectrum at each time step, with a specified wavelength resolution.

The intensity function \( I_{\nu,t,e}\) is determined by capturing the escaped particles and extracting the relevant information from them. As illustrated in Figure \ref{fig:snillu}, the fundamental concept is based on \cite{Huber2019}. The left panel shows a profile diagram perpendicular to the line of sight, where statistical methods are used to calculate averages, assuming the symmetry of the model employed in the simulation. On each outermost shell of the supernova, particles escape in various angular directions. We record the energy \( dE\) within each observer’s time bin (\( dt\)) and frequency bin (\( d\nu\)). By defining the angle \( \theta\) between the direction of the escaping particles and the radial direction from the supernova center to the outermost shell, we can determine the total intensity of parallel light in a given observation plane as follows:
\begin{equation}\label{eqintensity}
 I_{t,\nu,e} = \frac{1}{4\pi} \frac{dE}{dtd\nu}\;,
\end{equation}
where \( 4\pi \) is used to normalize to a unit sphere. Alternatively, this can be understood as rotating the observation plane around the sphere to account for all escaped particles, and then dividing by \( 4\pi \) to ensure the correct normalization. Furthermore, as described by the geometric delay in \cite{Lucy2005}, for escaped particles with different \( \theta \) values relative to the same observation time, the original emission time \( t \) will vary according to \( t = t_e - \frac{r \cos \theta}{c} \), where \( r \) is the radius of the outermost shell and \( c \) is the speed of light. Thus, by defining \( p = r \sin \theta \) in terms of \( \theta \), we can derive the emitted specific intensity at time \( t \), as shown in Equation \ref{eq:intensity}. Following the approach in \cite{Huber2019}, once we have the spectral density at emission, we can express the observed spectral flux from a supernova occupying a solid angle \( \Omega_0 \) as:
\begin{equation}\label{eq.a2}
 F_{\nu, \mathrm{o}}=\int_{\Omega_0} I_{\nu, \mathrm{o}} \cos \theta_{\mathrm{p}} \mathrm{d} \Omega\;,
\end{equation}
where \( I_{\nu, \mathrm{o}} \) is the specific intensity received by the observer. Since the scale of supernovae is typically much smaller than the angular diameter distance \( D_A \), we can make the transformation \( \theta_P = \frac{P}{D_A} \), where \( \cos \theta_P \approx 1 \).
As mentioned earlier, since the intensity we obtain corresponds to parallel light, we can transform the original solid angle as \( \mathrm{d} \Omega = \mathrm{d} \phi \, \mathrm{d} \theta_{\mathrm{p}} \) to \( \theta_{\mathrm{p}} = \frac{1}{D^2} \, \mathrm{d} \phi \, \mathrm{d} p \). Therefore, Equation \ref{eq.a2} can be rewritten as:
\begin{equation}\label{eqfluxo}
 F_{\nu, \mathrm{o}}=\frac{1}{D_{\mathrm{A}}^{2}} \int_{0}^{2 \pi} \mathrm{d} \phi \int_{0}^{p_{\mathrm{s}}} \mathrm{d} p p I_{\nu, \mathrm{o}}\;.
\end{equation}

\begin{figure}[H]
\centering
\includegraphics[width=0.7\textwidth, angle=0]{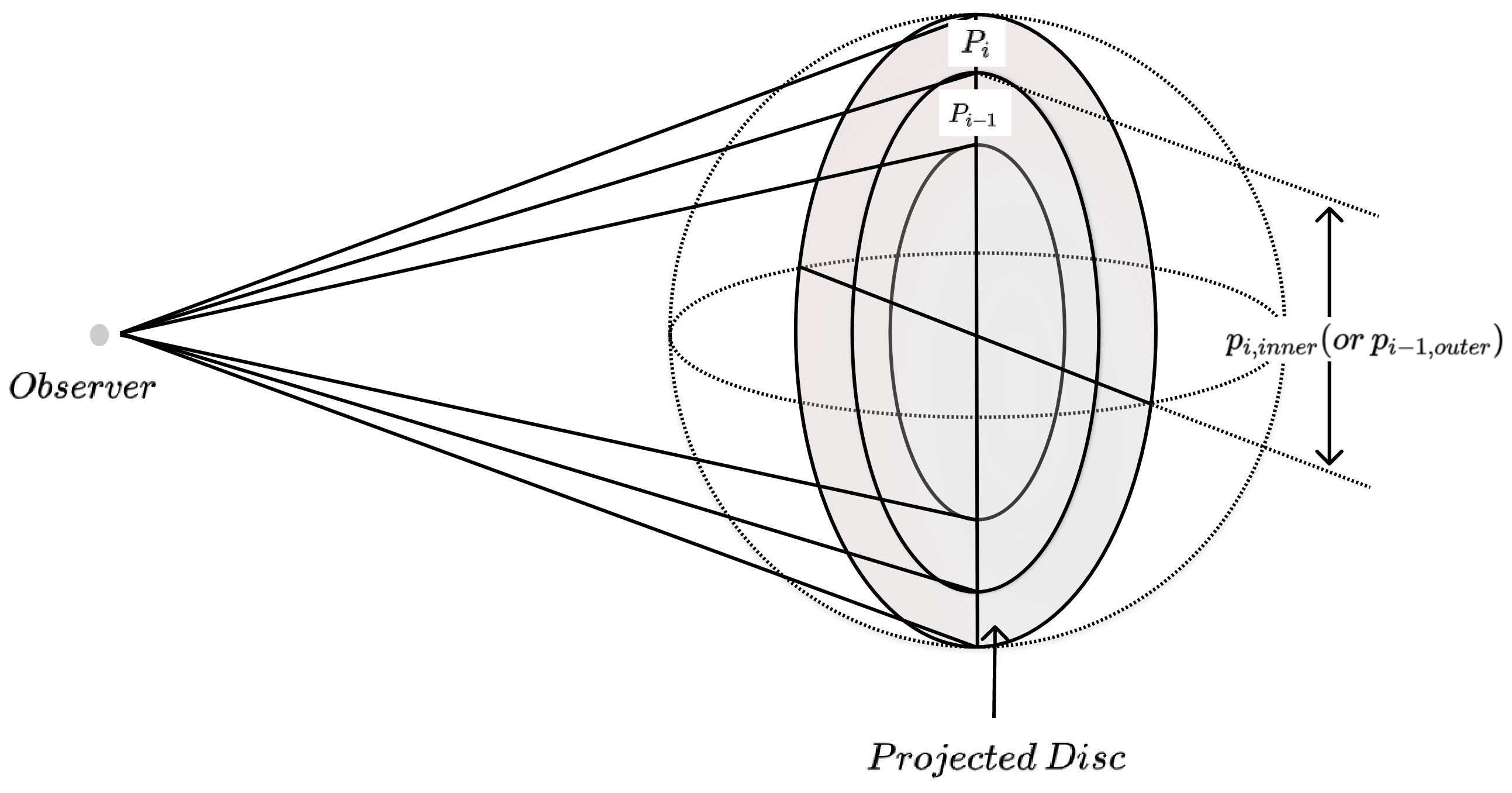}
\caption{The method of determining the photons from the supernova that reach the observer, after being projected into a two-dimensional plane, where \( P_i\) indicates the span of each annulus and \( p_i\) represents the radii at the inner and outer edges of each annulus.}
\label{fig:appen}
\end{figure}

According to Ref. \cite{Mihalas1984}, \( I_{\nu} / \nu^3 \) is Lorentz invariant, which implies that \( I_{\nu} \propto \nu^3 \). The relationship between the emitted frequency and the observed frequency can be determined through the redshift, \( \nu_{\mathrm{o}} = \frac{\nu_{\mathrm{e}}}{(1+z)} \), leading to the expression \( I_{\nu, \mathrm{o}} = I_{\nu, \mathrm{e}} (1+z)^{-3} \). After converting the angular diameter distance to the luminosity distance, we obtain:
\begin{eqnarray}\label{eqflux}
 F_{\nu, \mathrm{o}}&=&\frac{(1+z)}{D_{\mathrm{L}}^{2}} \int_{0}^{2 \pi} \mathrm{d} \phi \int_{0}^{p_{\mathrm{s}}} \mathrm{d} p p I_{\nu, \mathrm{e}}\nonumber \\
 &=&\frac{(1+z)}{D_{\mathrm{L}}^{2}} \sum_{i}\pi (P^2_{i,out}-P^2_{i,in})I_{\nu, \mathrm{e},P_i}\;,
\end{eqnarray}
where \( P_{i,out} \), \( P_{i,in} \), and \( P_i \) represent the outer and inner radii of the ring, respectively. After accounting for the microlensing magnification rate, as depicted in the right panel of Figure \ref{fig:appen}, the observed flux after magnification is given by:
\begin{eqnarray}
f_\nu &=& \mu F_{\nu, \mathrm{o}}=\frac{(1+z)}{D_{\mathrm{L}}^{2}} \int_{0}^{2 \pi} \mathrm{d} \phi \int_{0}^{p_{\mathrm{s}}} \mathrm{d} p p \mu I_{\nu, \mathrm{e}} \;,\nonumber\\
 &=& \frac{(1+z)}{D_{\mathrm{L}}^{2}} \sum_{i}\pi (P^2_{i,out}-P^2_{i,in})\mu_{Pi}I_{\nu, \mathrm{e},P_i}\;,\nonumber \\
 &=&\frac{1}{4\pi D_{\mathrm{L}}^{2}}\sum_{i}\mu_{Pi}(\frac{\mathrm{d} E}{\mathrm{d} t \mathrm{~d} (\nu/(1+z))})_{P_i} \;,
\end{eqnarray}
where \( \mu_{P_i} \) is the average microlensing magnification rate within the ring. This formula is consistent with the results derived in \cite{Goldstein2018, Huber2019}. To optimize storage and improve computational efficiency, we calculate the average magnification rate within each ring. The method for computing light curves follows the approach outlined in Ref. \cite{Bessell2012}:
\begin{equation}\label{eqmag}
 m_\mathrm{AB}=-2.5 \log \frac{\int f_\nu(\nu) S_x(\nu) d \nu / \nu}{\int S_x(\nu) d \nu / \nu}-48.60\;,
\end{equation}
where \( S_x(\nu) \) represents the transmission function for the \( g \), \( r \), and \( i \) bands corresponding to MOST.
\end{appendix}
\end{multicols}

%%%%%%%%%%%%%%%%%%%%%%%%%%%%%%%%%%%%%%%%

%\end{appendices}

\end{document}